\documentstyle[rotate,epsfig]{europhys}

\def\And{{\rm and\ }}

\def\stars{\bigskip\centerline{***}\medskip}

\newif\ifboo \boofalse

\begin{document}
\euro{}{}{}{}
\Date{9. April 1999}
\shorttitle{E. FREY {\sf et al.\/:} SCALING REGIMES AND CRITICAL
  DIMENSIONS IN THE KPZ PROBLEM}

\title{Scaling regimes and critical dimensions in the \\
  Kardar--Parisi--Zhang problem}
     
\author{E. Frey\inst{1,2}, U. C. T\"auber\inst{3} \And H. K.
  Janssen\inst{4}}

\institute{\inst{1} Lyman Laboratory of Physics, Harvard University,
  Cambridge, MA 02138, USA. \\ \inst{2} Institut f\"ur Theoretische
  Physik, Technische Universit\"at M\"unchen, James-Franck-Stra\ss e,
  85747 Garching, Germany. \inst{3} Department of Physics, Virginia
  Polytechnic Institute and State University, Blacksburg, Virginia
  24061-0435, USA. \inst{4}Institut f\"ur Theoretische Physik III,
  Heinrich-Heine-Universit\"at, 40225 D\"usseldorf, Germany.}

\rec{}{}

\pacs{\Pacs{05}{40$+$j}{Fluctuation phenomena, random processes, and
    Brownian motion} \Pacs{64}{60Ak}{Renormalization-group, fractal,
    and percolation studies of phase transitions}
  \Pacs{64}{60Ht}{Dynamic critical phenomena}}

\maketitle

\begin{abstract}
  We study the scaling regimes for the Kardar--Parisi--Zhang equation
  with noise correlator $R(q)\propto (1 + w \, q^{-2 \rho})$ in
  Fourier space, as a function of $\rho$ and the spatial dimension $d$.
  By means of a stochastic Cole--Hopf transformation, the critical and
  correction-to-scaling exponents at the roughening transition are
  determined to all orders in a $(d - d_c)$ expansion. We also argue
  that there is a intriguing possibility that the rough phases above
  and below the lower critical dimension $d_c = 2 (1 + \rho)$ are
  genuinely different which could lead to a re-interpretation of
  results in the literature.
\end{abstract}

The Kardar--Parisi--Zhang (KPZ) equation, which was originally
introduced to describe growth of rough
surfaces~\cite{kardar-parisi-zhang:86}, displays generic scale
invariance, as well as a non-equilibrium roughening transition
separating a smooth from a rough phase above the {\em lower} critical
dimension.  As a consequence of its mapping to the noisy Burgers
equation~\cite{forster-nelson-stephen:77}, to the statistical
mechanics of a directed polymer in a random
environment~\cite{huse-henley-fisher:85}, as well as to other
interesting equilibrium and non-equilibrium systems (for recent
reviews, see Refs.~\cite{halpin-healy-zhang:95,krug:97}), the KPZ
problem has emerged as one the fundamental theoretical models defining
possible universality classes for non-equilibrium scaling phenomena
and phase transitions.

In one dimension, the roughness and dynamic exponents, $\chi$ and $z$,
have long been determined exactly by means of the dynamic
renormalization group (RG), utilizing the symmetries of the problem
\cite{forster-nelson-stephen:77}.  Furthermore, it has been
demonstrated that the associated scaling functions can be computed to
high precision by means of the self-consistent one-loop or
mode-coupling approximation \cite{hwa-frey:91,frey-taeuber-hwa:96}.
For $d > d_c$, a two-loop RG calculation~\cite{frey-taeuber:94} indicated
that the critical behavior of the roughening transition might be
described by an exact set of exponents as suggested earlier on the
basis of scaling arguments~\cite{doty-kosterlitz:92}.  Using a
directed-polymer representation, L\"assig was able to demonstrate the
validity of this statement to all orders in a $(2 + \varepsilon)$
expansion \cite{laessig:95} (see also
Refs.~\cite{janssen_unpub:97,wiese:98}). In addition, these results
were evidence for a upper critical dimension $d_{\rm uc} = 4$ of the
roughening transition, and suggested that the ensuing strong-coupling
rough phase was not accessible within perturbation theory.  This is in
accord with the divergence of the coupling constant $g$ upon
approaching the lower critical dimension from
below~\cite{frey-taeuber:94}.

The scaling behavior in the strong-coupling rough phase above $d_c$
has been very controversial.  Based on very different assumptions and
analytic approaches, diverse values for the scaling exponents have
been postulated, see, e.g.,
Refs.~\cite{halpin-healy:90,stepanow:97,laessig:98}.  In addition,
some authors have claimed $d_{{\rm uc}} = 4$ to be the upper critical
dimension for the rough phase as well
\cite{moore-etal:95,laessig-kinzelbach:97,bhattacharjee:98}, where the
scaling exponents assume the values known in infinite dimensions
\cite{derrida-spohn:88,mezard-parisi:91}.  In contrast, numerical
studies observed merely continuously varying exponents as $d$ was
increased \cite{ala_nissila-hjelt-kosterlitz-venaelaeinen:93}.  In
addition, the validity of the continuum Langevin description has been
questioned in this regime, and a conceivable breakdown of universality
has been conjectured (see, e.g.,
Refs.~\cite{marsili-bray:96,newman-swift:97}).

In order to shed light on some of these open issues, we shall find it
useful to add a long-range power-law contribution to the usual
spatially local stochastic noise of the KPZ equation, as first
introduced by Medina {\em et~al.} \cite{medina-hwa-kardar-zhang:89}.
More generally, we investigate the Langevin equation,
\begin{equation}
        \partial_t s({\bf x},t) = D \, {\bf \nabla}^2 s({\bf x},t) 
        + {D \, g \over 2} \, \left[ {\bf \nabla} s({\bf x},t) \right]^2 
                + \zeta({\bf x},t) \ ,
\label{kpzeqn}
\end{equation}
with Gaussian noise characterized by zero mean and variance $\langle
\zeta({\bf x},t) \, \zeta({\bf x}^\prime,t^\prime) \rangle = 2 \,
R_0({\bf x}-{\bf x}^\prime) \, \delta(t-t^\prime)$, which we assume to
be {\em local} in time, but which may contain spatially long-range
power law contributions of the form $R_0({\bf x}-{\bf x}^\prime)
\propto |{\bf x} - {\bf x}^\prime|^{2 \rho - d}$.  Thus, typically we
have $R_0({\bf q}) = D \left( 1 + w \, |{\bf q}|^{- 2 \rho} \right)$
in Fourier space.  Notice that setting ${\bf u} = - {\bf \nabla} s$ in
Eq.~(\ref{kpzeqn}) leads to the noisy Burgers equation, with the usual
locally {\em conserved} noise for $w = 0$ or $\rho = 0$, but with {\em
  non-conserved} noise for $\rho = 1$ (termed model B in
Ref.~\cite{forster-nelson-stephen:77}). We find as the most important
effects of adding the long-range noise term: (i) the lower critical
dimension for the roughening transition is shifted {\em upwards} to
$d_c = 2 \left( 1 + \rho \right)$; (ii) above $d_c$, there are {\em
  two} subtly distinct regimes for the {\em smooth} phase,
characterized by different correction-to-scaling exponents; (iii)
below $d_c$, there are two distinct {\em rough} regimes, governed by
the short-range and long-range noise RG fixed points, respectively,
and separated by a line $\rho_*(d)$ in the $(\rho,d)$ plane.

In the following, above $d_c$, we derive an {\em exact} integral
equation for the noise correlation function by means of a stochastic
Cole--Hopf transformation.  Based on the ensuing minimally
renormalized RG flow functions, we compute critical exponents at the
roughening transition to {\em all} orders in a $(d - d_c)$ expansion,
and determine the {\em exact} scaling exponents in the smooth phase
above $d_c$.  We demonstrate that the strong-coupling rough phase
above $d_c$ is perturbationally inaccessible, but probably
characterized by a spatially local noise correlator.  Below the lower
critical dimension $d_c$, we determine the scaling exponents at the
long-range noise fixed point {\em exactly}, provided such a
non-trivial finite fixed point exists.  We discuss different
analytical RG approaches to such a strong-coupling fixed point, and
obtain an approximate expression for the separatrix $\rho_*(d)$.
Finally, we discuss whether the rough phases above and below $d_c$,
are ``continuously'' connected as a function of space dimension $d$
and the magnitude of the noise correlation exponent $\rho$. We shall
argue that there is a intriguing possibility that these two rough
phases are genuinely different.

\begin{figure}
  \centerline{\epsfxsize 0.6\columnwidth \epsfbox{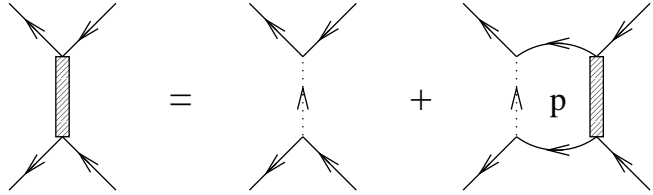}}
  \vspace{0.1cm}
  \caption{Bethe--Salpeter equation for the noise correlator.}
\label{fig:bethe_salpeter}
\end{figure}

A convenient starting point for a systematic analysis of a Langevin
equation like Eq.~(\ref{kpzeqn}) is its reformulation in terms of a
dynamic generating functional \cite{frey-taeuber:94},
\begin{eqnarray}
        {\cal J} [{\tilde s},s] &=& \int \! d^dx \int \! dt \; {\tilde s}({\bf x},t) 
        \biggl[ \partial_t s({\bf x},t) - D \, {\bf \nabla}^2 s({\bf x},t)
\label{dynfun} \\
        &&- {D \, g \over 2} \left[ {\bf \nabla} s({\bf x},t) \right]^2 
        - \int \! d^dx^\prime R_0({\bf x}-{\bf x}^\prime) \, 
                {\tilde s}({\bf x}^\prime,t) \biggr] \ , \nonumber 
\end{eqnarray}
which allows the calculation of expectation values by means of path integrals
with the statistical weight $\exp(-{\cal J}[{\tilde s},s])$. Upon directly
applying a Cole--Hopf transformation to the KPZ equation,
Eq.~(\ref{kpzeqn}), one obtains a diffusion equation subject to
multiplicative noise which is often interpreted as a directed polymer
in a random potential~\cite{huse-henley-fisher:85}. If we recast the
same idea in terms of the above dynamic functional the corresponding
stochastic Cole--Hopf transformation \cite{janssen_unpub:97} --- in
the appropriately discretized version of Eq.~(\ref{dynfun}) in the Ito
representation --- reads
\begin{eqnarray}
        n({\bf x},t) &=& {2 \over g} \, \exp \left\{ {g \over 2} \, 
        \bigl[ s({\bf x},t) + D \, R_0({\bf 0}) \, t \bigr] \right\} \ ,
\label{holktr} \\
        {\tilde n}({\bf x},t + \tau) &=& {\tilde s}({\bf x},t + \tau) 
        \exp \left\{ - {g \over 2} \bigl[ s({\bf x},t) + D \, R_0({\bf 0}) \, t
                \bigr] \right\} \, . \nonumber
\end{eqnarray}
This leads to a dynamic generating functional,
\begin{eqnarray}
        &&{\cal J} [{\tilde n},n] 
        = \int \! d^dx \int \! dt \biggl\{ {\tilde n}({\bf x},t) 
        \Bigl[ \partial_t n({\bf x},t) - D\, {\bf \nabla}^2 n({\bf x},t) \Bigr]
\label{holkdf} \\
        &&- {D \, g^2 \over 4} \int \! d^dx^\prime \, {\tilde n}({\bf x},t) \,
        n({\bf x},t) \, R_0({\bf x}-{\bf x}^{\prime}) \, 
        \tilde{n}({\bf x}^{\prime },t) \, n({\bf x}^{\prime },t) \biggr\} \ ,  
        \nonumber 
\end{eqnarray}
very reminiscent of the field theory for diffusion-limited pair
annihilation \cite{lee:94}. A remarkable feature of ${\cal J} [{\tilde
  n},n]$ is that there exist {\em no} loop diagrams contributing to a
renormalization of the diffusion propagator. Hence the dynamic
exponent is $z = 2$ whenever standard perturbation theory is
applicable.  This leaves us with the renormalization of the four-point
noise vertex, which is formally achieved to {\em all} orders in the
perturbation expansion via a Bethe-Salpeter equation, as graphically
depicted in Fig.~\ref{fig:bethe_salpeter}.  Analytically, this leads
to an integral equation for the renormalized noise correlator $R$,
\begin{eqnarray}
        R({\bf k},{\bf k}^\prime;\mu^2) = R_0({\bf k}-{\bf k}^\prime) 
        + {g^2 \over 4} \int \! {d^dp \over (2\pi)^d} \, 
        {R_0({\bf k}-{\bf p}) \over p^2 + \mu^2} \, R({\bf p},{\bf k}'; \mu^2) 
                \ ,
\label{betsal}
\end{eqnarray}
where $\mu^2 = i \omega / 2 D + q^2 / 4$, and ${\bf q}$ and $\omega$
are the momentum and frequency transfers from right to left in the
vertices of Fig.~\ref{fig:bethe_salpeter}, while ${\bf k}-{\bf
  k}^\prime$ denotes the momentum transfer from bottom to top.  The
standard perturbation expansion in terms of $g^2$ is equivalent to the
Neumann series for this Fredholm integral equation.  The {\em exact}
relation (\ref{betsal}) can be used to determine the scaling function
for the renormalized noise self-consistently
\cite{janssen-taeuber-frey:98}.

In this letter, we focus on the asymptotic scaling exponents.  Upon
inserting the noise correlator into the Bethe--Salpeter
equation (\ref{betsal}), one readily obtains the full renormalization
of the short-range noise strength $D$ within the minimal-subtraction
procedure, which here leads to a formal {\em double} expansion with
respect to $\varepsilon = d - 2$ {\em and} $\rho$.  On the other hand,
the non-analytic long-range contribution itself is not renormalized
perturbationally.  In terms of the renormalized counterparts $u$ and
$v$ for the bare couplings $g^2$ and $w \, g^2$, respectively, the
ensuing {\em exact} RG $\beta$ functions are actually precisely those
of the one-loop theory,
\begin{equation}
  \beta_u = \varepsilon \, u - (u + v)^2 / 2 \ , \quad
  \beta_v = (\varepsilon - 2 \rho) \, v \ .
\label{mnbeta}
\end{equation}
For $d > d_c = 2(1+\rho)$, Eq.~(\ref{mnbeta}) shows that $v \to 0$
asymptotically.  Therefore, the usual {\em short-range} noise KPZ
scenario applies
\cite{forster-nelson-stephen:77,kardar-parisi-zhang:86,%
  medina-hwa-kardar-zhang:89,frey-taeuber:94}.
 
In addition to the trivial fixed point $u_* = 0$ and the
strong-coupling fixed point $u_* = \infty$, there appears an {\em
  unstable} critical fixed point $(u_c,v_c) = (2 \, \varepsilon,0)$,
marking the location of a non-equilibrium {\em roughening transition}
above the {\em lower} critical dimension $d_c$.  At this
$O(\varepsilon)$ fixed point, the general scaling relation $\chi + z =
2$ is valid, and because of $z_c = 2$ we find the marginal roughness
exponent $\chi_c = 0$ \cite{doty-kosterlitz:92,laessig:95}.
Furthermore, the negative eigenvalue of the associated stability
matrix defines the crossover or inverse correlation length exponent
$\phi_c = \nu_c^{-1} = d - 2$, whereas its positive eigenvalue yields
the correction-to-scaling exponent $\omega_c = d - d_c$.  Assuming
that the Chayes--Fisher bound $\nu > 2/d$
\cite{chayes-chayes-fisher-spencer:86} applies for the crossover
length scale in this problem (equivalent to a directed polymer in a
random environment), one finds that this lower bound is reached in
$d_{{\rm uc}} = 4$ dimensions, which therefore constitutes the {\em
  upper} critical dimension for the roughening transition, beyond
which $\phi_c = 2$, $\nu_c = 1/2$ \cite{laessig:95,taeuber-frey:95}.

In the {\em smooth} phase, both couplings $u \to 0$ and $v \to 0$, and
a more careful analysis of the RG flow of the ratio $w = v / u$ is
required (details will be presented elsewhere
\cite{janssen-taeuber-frey:98}).  One finds that actually $w \to
\infty$, and thus the smooth phase, as opposed to the roughening
transition, is characterized by the {\em long-range} algebraic noise,
with roughness exponent $\chi_{{\rm sm}} = 1 + \rho - d / 2 \leq 0$,
and $z_{{\rm sm}} = 2$.  Furthermore, {\em two distinct} scaling
regimes emerge, distinguished by different correction-to-scaling
exponents: For $d_c < d < 2 (1 + 2 \rho)$, one finds $\omega_1 = d - 2
(1 + \rho)$ and $\omega_2 = 2 (1 + 2 \rho) - d$, whereas for $d \geq
2(1+2\rho)$, $\omega_1 = d - 2 (1 + 2 \rho)$ and $\omega_2 = 2 \rho$.
Precisely at $d = d_c$, i.e., for $\rho = (d - 2) / 2$, there appears
a fixed {\em line}, which is unstable for $w < 1$, and stable for $w >
1$.

Finally, the RG analysis also tells us that the rough phase emerging
above $u_c$ is a genuine {\em strong-coupling} phase in the sense that
it remains inaccessible to perturbative methods, even to {\em all}
orders in $\varepsilon$.  Numerical solutions of the flow equations
also suggest that $w \to 0$ in the rough phase; hence the algebraic
noise correlations appear to be {\em irrelevant} in the
strong-coupling regime.

Up to now we have restricted our discussion to dimensions larger than
the lower critical dimension, where there is a kinetic roughening
transition from a smooth to a rough phase. If one tries to extend the
above analysis to $d< d_c = 2(1+\rho)$, where there exists {\em no}
roughening transition, Eq.~(\ref{mnbeta}) implies that $v \to \infty$
at long length scales.  Thus, the {\em minimally} renormalized
perturbation theory based on the stochastic Cole--Hopf transformation
breaks down, and one must resort to other methods. Fortunately,
however, we can draw some important {\em exact} conclusions already
from the general structure of the field theory (\ref{dynfun}).  As a
consequence of (a) the underlying Galilean invariance, which fixes the
renormalization of $g$, (b) the fact that the non-analytic noise term
proportional to $D w$ cannot renormalize, and (c) the momentum
dependence of the non-linear vertex, there are merely {\em two}
independent renormalization constants to be determined, namely for the
renormalized fields, $s_R = Z^{1/2} \, s$ and the renormalized
diffusion constant, $D_R = Z_D D$.  Upon defining $\gamma = \mu
\partial_\mu \vert_0 \ln Z$ and $\zeta = \mu \partial_\mu \vert_0 \ln
Z_D$, the RG $\beta$ functions can be expressed as $\beta_u = (d - 2 -
\gamma - 2 \, \zeta) u$ and $\beta_v = (d - 2 - 2 \, \rho - 3 \,
\zeta) v$.  Then by solving the RG equation for the correlation
function near a stable RG fixed point, we furthermore identify $\chi =
(2 - d + \gamma_*) / 2$ and $z = 2 + \zeta_*$.  The existence of a
{\em non-zero, finite} RG fixed point $u_*$ then immediately leads to
the scaling relation $\chi + z = 2$ \cite{frey-taeuber:94}.
Similarly, at {\em any} long-range fixed point $0 < v_* < \infty$,
$\zeta_*$ is fixed, giving the {\em exact} values
\begin{equation}
        z_{{\rm lr}} = (4 + d - 2 \, \rho) / 3 \ , \quad
        \chi _{{\rm lr}} = (2 - d + 2 \, \rho) / 3
\label{lrexpo}
\end{equation}
for the scaling exponents at the long-range fixed point, {\em provided} 
$0 < u_* < \infty$ as well.

In general, these two scaling fixed points compete, and the
short-range fixed point must evidently be dominant, if $z_{{\rm sr}} <
z_{{\rm lr}}$ (and vice versa), which indeed implies $\beta_v > 0$ and
hence $v \to 0$.  In one dimension, and for purely local noise ($w =
0$), the non-linear reversible force term proportional to $g$ can be
shown to fulfill the detailed-balance conditions, guaranteeing that the
stationary probability distribution becomes ${\cal P}_{{\rm st}}
\propto \exp \left( -\int dx [\nabla s(x)]^2 /2 \right)$, as for the
linear equation.  In this Gaussian static theory, there can be no
field renormalization ($Z = 1$), hence $\zeta = 0$, leading to the
familiar one-dimensional KPZ short-range scaling exponents $\chi_{{\rm
    sr}} = 1 / 2$ and $z_{{\rm sr}} = 3 / 2$
\cite{forster-nelson-stephen:77,kardar-parisi-zhang:86}.  Comparing
the latter with the long-range dynamic exponent $z_{{\rm lr}} = (5 - 2
\rho) / 3$, we find that the short-range fixed point remains {\em
  stable}, provided $\rho < 1/4$ \cite{medina-hwa-kardar-zhang:89}.
After some controversy in the literature, this result has been
confirmed by simulations for the noisy Burgers equation
\cite{hayot-jayaprakash:96}.

Notice that a {\em minimal} renormalization scheme can never arrive at
the exact one-dimensional exponents $\chi_{{\rm sr}} = 1 / 2$ and
$z_{{\rm sr}} = 3 / 2$.  Instead, in order to address the rough phase
below $d_c$, one may devise a {\em non-minimal} renormalization
procedure at fixed dimension $d$ and $\rho$
\cite{frey-taeuber:94,janssen-taeuber-frey:98}.  Alternatively, one
may utilize the mapping to the Burgers equation and hence to driven
diffusive systems, for which a well-defined $(2 - d)$ expansion
exists.  Adding long-range correlated noise, this actually leads to
the identical stability condition $\rho < 1/4$ for the short-range
fixed point in $d = 1$ \cite{janssen-schmittmann:98}.  In the
long-range regime, the case $\rho = 1$, corresponding to the Burgers
equation with {\em non-conserved} noise, is accessible through an
$\varepsilon$ expansion below the upper critical dimension $d_{{\rm
    uc}} = 4$ of this model \cite{forster-nelson-stephen:77}.  The
dynamic exponent here is actually obtained to all orders in
$\varepsilon$, and reads $z_{{\rm lr}} = (2 + d) / 3$.

\begin{figure}
  \centerline{\epsfxsize 0.8\columnwidth \epsfbox{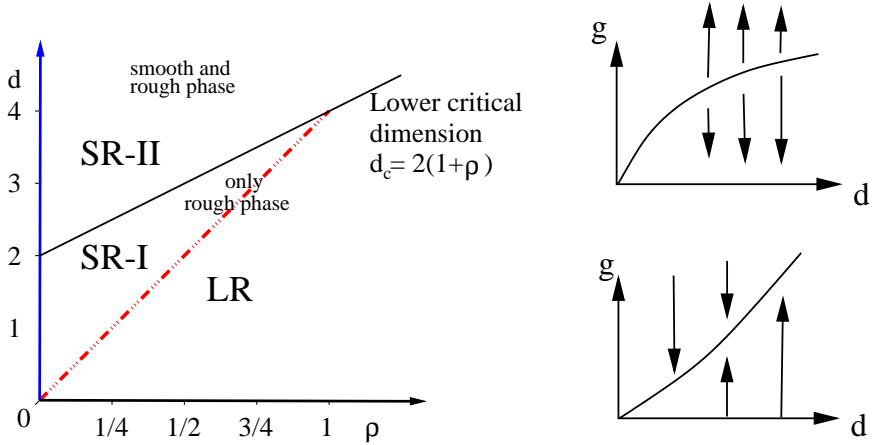}}
  \vspace{0.1cm}
  \caption{Scaling regimes (left) and sketches of the RG flow 
    of the coupling constant $g$ (right) above and below $d_c (\rho)$. 
   The dashed-dotted line indicates a approximate separatrix between
   the domain of attraction of the short-range and long-range noise
   fixed point.}
\label{fig:phase_diagram}
\end{figure}

In order to further discuss the implications of the above {\em exact}
results for the KPZ equation with long-range correlated noise it is
instructive to consider the scaling regimes in the $(d,\rho)$
landscape; see Fig.~\ref{fig:phase_diagram}. One should notice that
there are two {\em qualitatively} distinct regions separated by the
lower critical dimension $d_c(\rho) = 2 (1 + \rho)$.
For $d < d_c(\rho)$ there is only a rough phase and the short-range
(SR) and long-range (LR) noise fixed points compete. Above the lower
critical dimension there appears a phase transition from a smooth to a
rough phase and the RG analysis indicates that LR noise constitutes an
irrelevant perturbation in the rough phase.

{F}rom the above analysis we know the values of the scaling exponents in
the domain of attraction of the LR fixed point exactly.  Since the
scaling exponents at the SR fixed point are independent of $\rho$, the
only additional information needed to determine the SR exponents
is the separatrix $\rho_* (d)$ between the domains of attraction of
the LR and SR fixed points.  Of course, for the latter to be true one
has to require that the exponents are continuous at $\rho_* (d)$. This
has been shown explicitly for $d=1$, and seems very reasonable in
general. From the analysis in this letter we cannot determine the
exact form of the separatrix. Yet, we actually can locate some
points on this curve.  In $d = 1$, the short-range fixed point is
stable for $\rho < \rho_* = 1/4$; for $\rho = 1$ (the Burgers equation
with non-conserved noise), and below $d_{{\rm uc}} = 4$, there is only
the long-range regime.  If we assume that the separatrix $\rho_*(d)$
between the short-range and long-range regimes extends up to four
dimensions, then it definitely contains the points $(1,4)$, $(1,1/4)$,
and probably also $(0,0)$ in the $(\rho,d)$ plane.  A simple linear
interpolation yields the function $\rho_*(d) \approx d / 4$ (dashed
line in Fig.~\ref{fig:phase_diagram}); recent computer simulations in
fact found $\rho_*(2) \approx 1/2$ \cite{li:97}.  Inserting the linear
interpolation approximation (which we do not expect to be exact), 
$\rho_*(d) \approx d/4$, into Eq.~(\ref{lrexpo}) yields
\begin{equation}
        z_{{\rm sr}} \approx (8 + d) / 6 \ , \quad
        \chi _{{\rm sr}} \approx (4 - d) / 6 \ .
\label{srexpo}
\end{equation}
These values remarkably coincide with Halpin-Healy's results obtained
in a functional RG \cite{halpin-healy:90} and also with a more recent
perturbative mode-coupling study by
Bhattacharjee~\cite{bhattacharjee:98}.

There are now two possible scenarios. The first one is that all the
scaling exponents for the strong-coupling phase are continuous over
the whole $(\rho,d)$ plane and in particular upon crossing the line
marking the lower critical dimension. This would necessarily imply (if
not prove) that $d_{\rm uc} = 4$ is the {\em upper} critical dimension
for the rough phase in agreement with some recent
speculations~\cite{laessig-kinzelbach:97} but disagreement with
computer
simulations~\cite{ala_nissila-hjelt-kosterlitz-venaelaeinen:93}.
However, one should probably expect to encounter singularities as the
lower critical dimension $d_c(\rho)$ is crossed.  A second more
intriguing scenario is therefore that there are two fundamentally
different rough phases, one below and one above the lower critical
dimension $d_c (\rho) = 2 (1 + \rho)$, which we call type-I (SR-I) and
type-II (SR-II). This would then allow for the scaling exponents to be
discontinuous at $d_c (\rho)$. Below $d_c (\rho)$ one would have
scaling exponents close to those obtained from the linear
interpolation of the separatrix in Eq.~(\ref{srexpo}). In particular
the upper critical dimension of the type-I rough phase is $d_{uc} =4$.
Above the lower critical dimension $d_c (\rho)$ one would have a
different set of exponents (e.g.  those given by the numerical
simulations) and the upper critical dimension must not necessarily be
equal to $4$ or any other finite value. At present there is proof for
neither of the above scenarios but the following observations are
quite indicative. First, from the RG analysis exploiting the
Cole--Hopf transformation we have learned that the strong-coupling
phase above $d_c(\rho)$ is not accessible by perturbation theory even
to infinite order. On the other hand the rough phase at $d=1$ is
accessible by standard perturbation theory using a mapping of the KPZ
equation to a driven diffusion model~\cite{janssen-schmittmann:98}.
Second, for $\rho=0$ an explicit two-loop
calculation~\cite{frey-taeuber:94} shows that the fixed point value of
the coupling constant $g$ approaches infinity as the lower critical
dimension approaches $2$ from below.  Third, this scenario would
provide a coherent picture for most of the available numerical and
analytical results for the KPZ equation. Some of the analytical
approaches (e.g. functional RG and mode-coupling theory) are
self-consistent theories and hence necessarily start out with
correlated noise. This in effect shifts the lower critical dimension
upwards, and it is well possible that this automatically constrains
the results to the SR-I phase as opposed to the SR-II phase these
studies were supposedly aiming at. It is difficult to judge on other
analytical methods and how they fit into the scheme discussed here.
Lastly, there have been suggestions for a breakdown of a continuum
description, and perhaps even universality, in the rough phase
\cite{marsili-bray:96,newman-swift:97}.  Physically, in a
microscopically rough regime, the underlying lattice as well as
details of the dynamic rules may well be important; in addition, one
might question the existence of a well-defined short-distance
expansion \cite{laessig:98} in this phase.  Clearly, these issues
require further clarification through approaches that extend beyond
the continuum equation, Eq.~(\ref{kpzeqn}), and equivalent field
theory methods.

\stars

We benefited from discussions with J.L.\ Cardy, T.\ Halpin-Healy, and
K.J.\ Wiese.  E.F. and U.C.T. are grateful for support from the DFG
through Heisenberg and habilitation fellowships Fr 850/3-1 and Ta
177/2-1,2.  H.K.J. acknowledges support from the SFB 237 (``Unordnung
und gro\ss e Fluktuationen''), funded by the DFG.

\end{document}